\documentclass[letterpaper, 10 pt, journal, twoside]{IEEEtran}

\usepackage{graphicx}
\usepackage{booktabs}
\usepackage{amsmath}
\usepackage{amssymb}
\usepackage{subfig}
\usepackage{mathtools}
\newtheorem{theorem}{\textbf{Theorem}}
\newtheorem{assume}{\textbf{Assumption}}
\newtheorem{remark}{\textbf{Remark}}

\usepackage{xcolor}
\usepackage[normalem]{ulem}
\usepackage{epstopdf}
\usepackage{hyperref}
\usepackage[mathscr]{euscript}
\newcommand{\dif}{\mathop{}\!\mathrm{d}}
\usepackage{amsfonts, xargs, tensor, units, cite, stmaryrd, mathrsfs, algpseudocode, comment}
\definecolor{orange}{rgb}{1.0, 0.49, 0.0}

\newcommand{\R}{\mathbb{R}}
\DeclareMathOperator{\diag}{diag}

\makeatletter
\@ifundefined{proof}{%
}{}
\makeatother

\title{Concurrent Learning-Based Inertial Parameter Estimation and Control of Quadrotors
}

\author{Sandesh Thapa$^{1}$ and He Bai$^{2}$
\thanks{$^{1}$S. Thapa is with the ETAIC Lab, University of Texas at Arlington,
        Arlington, TX 76019, USA, {\tt\small sandesh.thapa@uta.edu}}.
\thanks{$^{2}$H. Bai is with the School of Mechanical \& Aerospace Engineering,
      Oklahoma State University, Stillwater, OK 74078, USA. {\tt\small he.bai@okstate.edu}}
}

\newcommand{\RD}[1]{{\color{red}#1}}
\begin{document}

\maketitle
\thispagestyle{empty}
\pagestyle{empty}

\begin{abstract}
In this letter, we consider trajectory tracking control and inertial parameter estimation of a quadrotor UAV whose center of mass (CoM) does not coincide with its geometric center. Onboard payloads like an IMU, camera, or battery are not necessarily placed at its geometric center. Moreover, during flight these sensors might move because of external disturbances and cause a CoM offset.  Thus, these events and equipment introduce an offset and cause an additional moment about the CoM. We develop an integral concurrent learning (ICL)-based geometric controller that estimates the mass, the CoM offset, and the principal moments of inertia online, without an accelerometer or numerical derivatives, and tracks a given trajectory corrected by the offset estimated online. We prove that the estimated parameters converge to their true values under a finite excitation condition. We validate the effectiveness of the proposed algorithm using simulation examples.
\end{abstract}


\section{INTRODUCTION}

\IEEEPARstart{Q}{uadrotors} are a popular research platform because of their agility and their ability to fit in tight spaces~\cite{lee2010arxiv},\cite{wu2025l1quad}. These systems have also been increasingly used in delivery and rescue operations~\cite{ruggiero2018aerial},\cite{goodman2026observer}. Such dynamic actions, along with the movement of the payload and battery in flight because of external disturbances, often shift the center of mass (CoM) away from the origin of its body frame, which is generally modeled as the geometric center for quadrotors~\cite{palunko2011adaptive,wuest2019online}. This offset creates a gravity moment and further makes the inertia matrix non-diagonal because of the coupling of the moment~\cite{palunko2011adaptive}.

Adaptive, geometric, and integral controllers have been actively studied to address such problems (see, e.g.,~\cite{wu2025l1quad}, \cite{perez2019adaptive}). These algorithms handle an unknown mass~\cite{perez2019adaptive,thapa2019concurrent},  inertia~\cite{perez2019adaptive,yang2022parameter, romero2025simultaneous}, or constant disturbances~\cite{goodarzi2013geometric}. The inertia matrix is often modeled as a diagonal matrix, and the offset is not estimated except in a few cases~\cite{palunko2011adaptive,wuest2019online},\cite{sharma2022tracking}. Palunko and Fierro~\cite{palunko2011adaptive} design an adaptive control algorithm that learns the unknown offset dynamically and adds the induced forces and moments created by the offset to a PD controller to track the attitude and position. Recently, Yang and Cheng~\cite{yang2022parameter} have designed an ICL-based estimator for a unknown inertia and mass for multirotors. However,~\cite{yang2022parameter} assumes that the CoM coincides with the geometric center. With an offset, the estimated inertia is biased. An integral term, as in~\cite{goodarzi2013geometric}, helps in correcting such a constant offset, but those methods identify neither the offset nor the inertia. Similarly,  Sharma and Kar~\cite{sharma2021adaptive} develop an adaptive geometric controller for a drone with an offset and a diagonal inertia and show that the errors are bounded.  Moreover, W\"uest et al.~\cite{wuest2019online} design an unscented Kalman filter (UKF) to estimate all the inertial parameters of a drone using sensors. Similar filter-based methods have also been developed in~\cite{kwon2023real,bohm2021filter},\cite{eschmann2024data}. Recently, the authors of~\cite{alrawashdeh2024inflight} estimate the mass and inertia using accelerometers. In contrast, our method requires neither accelerometers nor numerical derivatives.

Integral concurrent learning has been used effectively to estimate inertial parameters~\cite{parikhintegral,kamalapurkar2017concurrent,yang2022parameter,thapa2019concurrent},\cite{lee2024concurrent,obuz2024adaptive,hart2025system,shen2024composite}.
For payload transport, there exist models where the offset is considered, but it is assumed known~\cite{zeng2019geometric}.
In this paper, we apply ICL-based learning for a single quadrotor with a CoM offset, drawing from our previous work on ICL-based cooperative manipulation. 


The contribution of this paper is the development and stability analysis of an ICL-based geometric controller for a quadrotor with an unknown CoM offset, and the demonstration of its benefit, i.e., convergence of the mass, the CoM offset, and the inertia matrix without requiring a persistently exciting (PE) trajectory, an accelerometer, or numerical derivatives. Building on the geometric controller of~\cite{lee2010arxiv}, the desired CoM trajectory is tracked through the CoM position estimated online, and we prove local exponential convergence of the tracking and parameter estimation errors under a finite excitation condition. Parameter convergence typically leads to improved transient performance~\cite{yucelen2014improving}. Our algorithm includes (i) an offset-aware rotational regressor, which is obtained by integrating the rotational dynamics, and (ii) staged excitation steps, where the mass and offset can be learned successfully during hover and the inertia with x-y movement with rolling and pitching moments. We consider a constant CoM offset to illustrate the ICL-based control algorithm. Extension to dynamic changes in CoM will be considered in future work.

The rest of the paper is organized as follows. We formulate the dynamic model in Section~\ref{sec:pf}. In Section~\ref{sec:control_design}, we develop our estimation and control algorithm. We present the stability analysis in Section~\ref{sec:stability}. Simulation results and conclusions are presented in Sections~\ref{sec:sim} and~\ref{sec:conc}, respectively.

\section{Problem Formulation} \label{sec:pf}
\begin{figure}[!h]
\centering
\includegraphics[width=0.75\columnwidth]{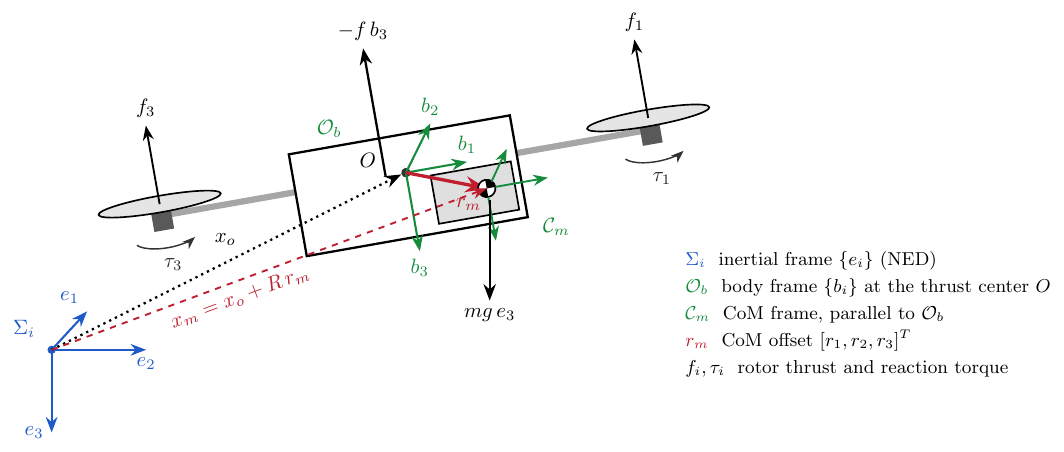}
\caption{Quadrotor with the inertial frame $\Sigma_i$ (north-east-down) and its attached body frame $\mathcal O_b$ at the geometric center $O$ and {$\mathcal C_m$} at the {CoM}  with an unknown offset {$r_m$}.}
\label{fig:quad}
\end{figure}
\subsection{Kinematics}
Let $\Sigma_i=\{e_1,e_2,e_3\}$ be the inertial frame with $e_3$ pointing down, and $\mathcal O_b=\{b_1,b_2,b_3\}$ be the body frame attached at geometric center $O$ (Fig.~\ref{fig:quad}). Let $x_o\in\R^3$ and $v_o\in\R^3$ be the position and velocity of $O$, $R\in SO(3)$ the attitude, and $\Omega\in\R^3$ the body angular velocity, $\widehat a\,b=a\times b$, and $(\cdot)^\vee$ is the inverse of $\widehat{\cdot}$. For $a=[a_1\ a_2\ a_3]^T$, the hat operator gives the skew-symmetric matrix
\begin{equation}\label{eq:hat}
\widehat a=\begin{bmatrix}0&-a_3&a_2\\ a_3&0&-a_1\\ -a_2&a_1&0\end{bmatrix},
\end{equation}
 $\widehat a^{\,T}=-\widehat a$. The kinematics of the body frame are
\begin{align}
\dot x_o&=v_o, & \dot R&=R\,\widehat\Omega .\label{eq:kinO}
\end{align}
Let $r_m=[r_1,r_2,r_3]^T$ be the constant position of the CoM relative to $O$, expressed in $\mathcal O_b$. The position and velocity of the CoM are
\begin{align}
x_m&=x_o+R\,r_m, \label{eq:xm}\\
v_m&=\dot x_m=v_o+R\,\widehat\Omega\,r_m, \label{eq:vm}
\end{align}
since $\dot r_m=0$ and $\dot R=R\widehat\Omega$.

\subsection{Dynamics}
Let $m$ be the mass and $J_m=\diag(J_{11},J_{22},J_{33})>0$ be the inertia matrix about the CoM. The translational and rotational dynamics are given by
\begin{align}
\dot x_m&=v_m,\label{eq:dyn1}\\
m\,\dot v_m&=mg\,e_3-fRe_3,\label{eq:dyn2}\\
\dot R&=R\,\widehat\Omega,\label{eq:dyn3}\\
J_m\dot\Omega+\Omega\times J_m\Omega&=M+r_m\times fe_3.\label{eq:dyn4}
\end{align}

The rotor thrusts $f_i$, $i=1,\dots,4$, act along $-b_3$ and generate the total thrust $f$ and the moment $M := [M_1, M_2, M_3]^T$ about $O$. Because of that, we define the translational dynamics about the CoM. We model the term $ r_m \times f e_3$ as the additional moment induced about the CoM by the thrust vector acting at $O$.

Furthermore, differentiating~\eqref{eq:vm} with $\dot r_m=0$, we get
\begin{equation}\label{eq:vmdot}
\dot v_m=\dot v_o+\frac{\mathrm d}{\mathrm dt}\big(R\widehat\Omega\big)\,r_m ,
\end{equation}
where, by the product rule and $\dot R=R\widehat\Omega$,
\begin{equation}\label{eq:dRW}
\frac{\mathrm d}{\mathrm dt}\big(R\widehat\Omega\big):=\dot R\,\widehat\Omega+R\,\widehat{\dot\Omega}
=R\big(\widehat\Omega\widehat\Omega+\widehat{\dot\Omega}\big).
\end{equation}

\subsection{Parameters and Estimation Errors}
\begin{remark}\label{rem:planar}
The induced moment term, $r_m\times fe_3$ in~\eqref{eq:dyn4}, does not contain $r_3$. Further analyzing~\eqref{eq:vmdot}, we can observe that $r_3$ enters the translational dynamics through the term $\tfrac{\mathrm d}{\mathrm dt}(R\widehat\Omega)r_m$. Hence, for the rest of the paper, we model $r_3 = 0$ as in~\cite{kwon2023real}. Further algorithms to identify $r_3$ for dynamic 3D trajectories and payload transport will be explored in future work.
Also note that even though the inertia matrix is not diagonal, its off-diagonal terms are negligible. Hence, we use a diagonal inertia matrix $J_m$~\cite{wuest2019online,yang2022parameter}.
\end{remark}

We define our parameter states as
\begin{equation}\label{eq:theta}
\theta_r:=[r_1,\ r_2,\ J_{11},\ J_{22},\ J_{33}]^T\in\R^5,\qquad m\in\R_{>0},
\end{equation}
with $\theta=[\theta_r^T,\ m]^T\in\R^{6}$ and $r_m=[r_1,\ r_2,\ 0]^T$. We define the estimation errors as
\begin{equation}\label{eq:esterr}
\tilde\theta_r:=\theta_r-\hat\theta_r,\qquad \tilde m:=m-\hat m,\qquad \tilde r:=r_m-\hat r,
\end{equation}
where $\hat r=[\hat r_1,\ \hat r_2,\ 0]^T$, and $\tilde\theta=[\tilde\theta_r^T,\ \tilde m]^T$. The error of the true CoM is further given by
\begin{equation}\label{eq:trueerr}
x_m-x_d=e_x+R\,\tilde r .
\end{equation}

Hence, the true CoM tracks $x_d$ only when both the position error $e_x \to 0$ and the offset estimation error $\tilde{r} \to 0$.  

\subsection{Tracking Errors}
From rigid-body kinematics, we get
\begin{equation}\label{eq:xhat}
\begin{aligned}
\hat x_m&=x_o+R\,\hat r,\\
\hat v_m&=v_o+R\,\widehat\Omega\,\hat r .
\end{aligned}
\end{equation}
We next define our tracking errors for a given trajectory $x_d(t)$ and a desired heading direction $b_{1d}(t)$,
\begin{equation}\label{eq:errors}
\begin{aligned}
e_x&=\hat x_m-x_d, & e_v&=\hat v_m-\dot x_d,
\end{aligned}
\end{equation}
where $R_d\in SO(3)$ is the desired attitude, and $\Omega_d$ is the desired angular velocity. The error on $SO(3)$ is given by~\cite{lee2010arxiv}
\begin{equation}\label{eq:Psi}
\Psi(R,R_d)=\tfrac12\operatorname{tr}\big(I_3-R_d^TR\big),
\end{equation}
which is locally positive definite about $R=R_d$ when the rotation angle between $R$ and $R_d$ is less than $180^\circ$, i.e., in the sublevel set $\{\Psi<2\}$. Its derivative gives the attitude error
\begin{equation}\label{eq:eR}
e_R=\tfrac12\big(R_d^TR-R^TR_d\big)^\vee,
\end{equation}
and the angular velocity error is $e_\Omega=\Omega-R^TR_d\Omega_d$.
\subsection{Control Objective}
Our control objective is to design the thrust $f$, the moment $M$, and the update laws for $\hat\theta_r$ and $\hat m$, for a given smooth desired trajectory $x_d(t)$ and heading $b_{1d}(t)$, such that all signals remain bounded and, under finite excitation, $(e_x,e_v,e_R,e_\Omega,\tilde\theta)\to0$. By~\eqref{eq:trueerr}, the true CoM then converges to the desired trajectory and $\hat m\to m$, $\hat r\to r_m$, $\hat J_m\to J_m$.

\section{Control Design}\label{sec:control_design}
\begin{figure}[!h]
\centering
\includegraphics[width=0.5\columnwidth]{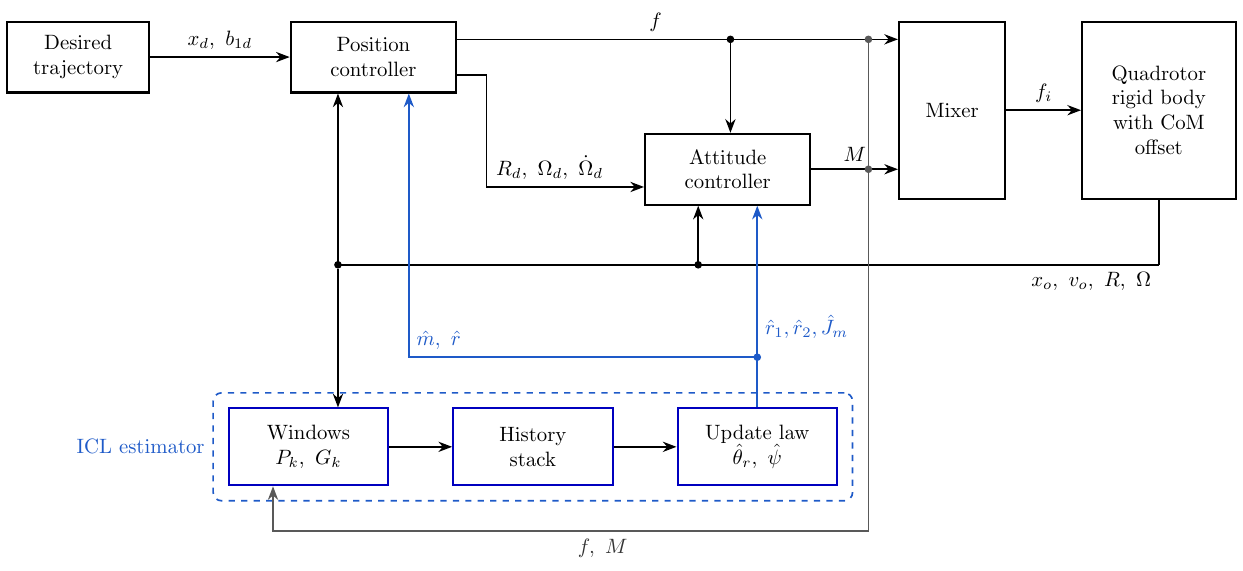}
\caption{Control architecture. }
\label{fig:arch}
\end{figure}
Figure~\ref{fig:arch} shows the cascaded geometric control architecture with the ICL-based estimator.

We are now ready to state our control and estimation laws.
\subsection{Position Controller}
For the control input $f$, we design our controller as:
\begin{equation}\label{eq:Fd}
F_d=-k_xe_x-k_ve_v+\hat m\,(\ddot x_d-ge_3),
\end{equation}
where $k_x,k_v\in\R_{>0}$. The thrust is given by
\begin{equation}\label{eq:f}
f=-F_d^TRe_3 .
\end{equation}
We compute the desired attitude as $R_d=[b_{2d}\times b_{3d},\ b_{2d},\ b_{3d}]$~\cite{lee2010arxiv}:
\begin{equation}\label{eq:Rd}
b_{3d}=-\frac{F_d}{\|F_d\|},\qquad b_{2d}=\frac{b_{3d}\times b_{1d}}{\|b_{3d}\times b_{1d}\|}.
\end{equation}

\subsection{Attitude Controller}
For the moment input $M$, our control law takes the following form,
\begin{equation}\label{eq:M}
\begin{split}
M&=-k_Re_R-k_\Omega e_\Omega+\hat J_m\big(R^TR_d\dot\Omega_d-\widehat\Omega R^TR_d\Omega_d\big)\\
&\quad+\Omega\times\hat J_m\Omega-\hat r\times fe_3.
\end{split}
\end{equation}

Equation~\eqref{eq:M} can be rewritten as
\begin{equation}\label{eq:Mreg}
\begin{aligned}
M&=-k_Re_R-k_\Omega e_\Omega+\hat J_m\big(R^TR_d\dot\Omega_d-\widehat\Omega R^TR_d\Omega_d\big)\\
&\quad+\Omega\times\hat J_m\Omega-\hat r\times fe_3\\
&=-k_Re_R-k_\Omega e_\Omega\\
&\quad+Y_r\big(R^TR_d\dot\Omega_d-\widehat\Omega R^TR_d\Omega_d,\ \Omega,\ f\big)\,\hat\theta_r ,
\end{aligned}
\end{equation}
where $k_R,k_\Omega\in\R_{>0}$, $\hat J_m$ is the estimate of $J_m$, and $Y_r$ is the regressor in~\eqref{eq:rotreg}--\eqref{eq:YfYJ} of Section~\ref{sec:Update_inertial}. For computing $\Omega_d$ and $\dot \Omega_d$, we refer the reader to~\cite{lee2010arxiv}.

\subsection{Inertial Parameter Estimation}\label{sec:Update_inertial}

\paragraph{Mass Estimation}
Substituting~\eqref{eq:vmdot} into~\eqref{eq:dyn2}, we rewrite the translational dynamics as
\begin{equation}\label{eq:transreg}
-fRe_3=\underbrace{\Big(\dot v_o-ge_3+\tfrac{\mathrm d}{\mathrm dt}\big(R\widehat\Omega\big)\,r_m\Big)}_{Y_t(r_m)}\,m,
\end{equation}
which is linear in $m$. Integrating~\eqref{eq:transreg} over the interval $[t-\tau,t]$ yields
\begin{equation}\label{eq:iclt}
\underbrace{-\int_{t-\tau}^{t}fRe_3\dif\gamma}_{P_t(t)}=\Big([v_o]-g\tau e_3+[R\widehat\Omega]\,r_m\Big)\,m ,
\end{equation}
 The offset enters~\eqref{eq:iclt} only through $[R\widehat\Omega]$, which vanishes if the vehicle does not rotate. Thus, the mass can be easily identified during hover, using the following regressor
\begin{equation}\label{eq:Gt}
G_t(t)=[v_o]-g\tau e_3\in\R^3.
\end{equation}
Furthermore, in hover, $G_t=-g\tau e_3$ and~\eqref{eq:iclt} simplifies to $\int f\dif\gamma=mg\tau$. With a history stack $\{(P_{k,t},G_{k,t})\}_{k=1}^{N_t}$, full rank if $0<\underline c_t<\mathscr G_t$, $\mathscr G_t:=\sum_{k=1}^{N_t}G_{k,t}^TG_{k,t}\in\R$, the concurrent learning update law for the mass is given by,
\begin{equation}\label{eq:clm}
\dot{\hat m}=k_t\Gamma_t\sum_{k=1}^{N_t}G_{k,t}^T\big(P_{k,t}-G_{k,t}\hat m\big),
\end{equation}
where $k_t\in\R_{>0}$ and the scalar gain $\Gamma_t>0$ is updated using
\begin{equation}\label{eq:Gamma_t}
\dot\Gamma_t=\beta_t\Gamma_t-k_t\Gamma_t\frac{\mathscr G_t}{1+\alpha_t\mathscr G_t} \Gamma_t,
\end{equation}
in which $\alpha_t,\beta_t\in\R_{>0}$ and $\Gamma_t(0)>0$.

\paragraph{Inertia and Planar Offset Estimation}
Rewriting the rotational dynamics~\eqref{eq:dyn4}, we get
\begin{equation}\label{eq:Mfull}
M=J_m\dot\Omega+\Omega\times J_m\Omega-r_m\times fe_3 ,
\end{equation}
Since $\Omega=[\Omega_1\ \Omega_2\ \Omega_3]^T$, $\dot\Omega=[\dot\Omega_1\ \dot\Omega_2\ \dot\Omega_3]^T$ and $J_m=\diag(J_{11},J_{22},J_{33})$,  we can re-write each term as, 
\begin{equation}\label{eq:JOmdot}
J_m\dot\Omega=\begin{bmatrix}J_{11}\dot\Omega_1\\ J_{22}\dot\Omega_2\\ J_{33}\dot\Omega_3\end{bmatrix}
=\begin{bmatrix}\dot\Omega_1&0&0\\ 0&\dot\Omega_2&0\\ 0&0&\dot\Omega_3\end{bmatrix}
\begin{bmatrix}J_{11}\\ J_{22}\\ J_{33}\end{bmatrix}.
\end{equation}
For $\Omega\times J_m\Omega$, $J_m\Omega=[J_{11}\Omega_1\ \ J_{22}\Omega_2\ \ J_{33}\Omega_3]^T$, and using the cross product and hat operator, 
\begin{equation}\label{eq:OmJOm}
\begin{aligned}
\Omega\times J_m\Omega&=\widehat\Omega J_m\Omega\\
&=\begin{bmatrix}0&-\Omega_3&\Omega_2\\ \Omega_3&0&-\Omega_1\\ -\Omega_2&\Omega_1&0\end{bmatrix}
\begin{bmatrix}J_{11}\Omega_1\\ J_{22}\Omega_2\\ J_{33}\Omega_3\end{bmatrix}\\
&=\begin{bmatrix}-J_{22}\Omega_2\Omega_3+J_{33}\Omega_2\Omega_3\\ J_{11}\Omega_1\Omega_3-J_{33}\Omega_1\Omega_3\\ -J_{11}\Omega_1\Omega_2+J_{22}\Omega_1\Omega_2\end{bmatrix}\\
&=\begin{bmatrix}0&-\Omega_2\Omega_3&\Omega_2\Omega_3\\ \Omega_1\Omega_3&0&-\Omega_1\Omega_3\\ -\Omega_1\Omega_2&\Omega_1\Omega_2&0\end{bmatrix}
\begin{bmatrix}J_{11}\\ J_{22}\\ J_{33}\end{bmatrix}.
\end{aligned}
\end{equation}
For $-r_m\times fe_3$, we have
\begin{equation}\label{eq:rfe3}
\begin{aligned}
-r_m\times fe_3&=-\widehat r_m fe_3
=-\begin{bmatrix}0&-r_3&r_2\\ r_3&0&-r_1\\ -r_2&r_1&0\end{bmatrix}
\begin{bmatrix}0\\ 0\\ f\end{bmatrix}\\
&=\begin{bmatrix}-fr_2\\ fr_1\\ 0\end{bmatrix}\\
&=f\begin{bmatrix}0&-1\\ 1&0\\ 0&0\end{bmatrix}
\begin{bmatrix}r_1\\ r_2\end{bmatrix},
\end{aligned}
\end{equation}
we can further notice that $r_3$ does not appear in the rotational dynamics, since the thrust acts along $e_3$.

Substituting~\eqref{eq:JOmdot}--\eqref{eq:rfe3} into~\eqref{eq:Mfull}, the rotational regressor takes the following form,
\begin{equation}\label{eq:rotreg}
M=\underbrace{\big[\,Y_f\ \ Y_J\,\big]}_{Y_r(\dot\Omega,\Omega,f)}\theta_r,\quad
\theta_r=[r_1,\ r_2,\ J_{11},\ J_{22},\ J_{33}]^T,
\end{equation}
where
\begin{equation}\label{eq:YfYJ}
\begin{aligned}
Y_f&=f\,Y_{fm},\qquad Y_{fm}=\begin{bmatrix}0&-1\\ 1&0\\ 0&0\end{bmatrix},\\
Y_J&=\begin{bmatrix}\dot\Omega_1&-\Omega_2\Omega_3&\Omega_2\Omega_3\\ \Omega_1\Omega_3&\dot\Omega_2&-\Omega_1\Omega_3\\ -\Omega_1\Omega_2&\Omega_1\Omega_2&\dot\Omega_3\end{bmatrix},
\end{aligned}
\end{equation}

Integrating~\eqref{eq:rotreg} over the interval $[t-\tau,t]$ for some constant $\tau\in\R_{>0}$ yields
\begin{equation}\label{eq:iclr}
\underbrace{\int_{t-\tau}^{t}M(\gamma)\dif\gamma}_{P_r(t)}
=\underbrace{\int_{t-\tau}^{t}Y_r\big(\dot\Omega(\gamma),\Omega(\gamma),f(\gamma)\big)\dif\gamma}_{G_r(t)}\ \theta_r ,
\end{equation}
where, using~\eqref{eq:YfYJ}, $G_r\in\R^{3\times5}$ is given by
\begin{equation}\label{eq:Gr}
G_r(t)=\big[\,G_f(t)\ \ G_J(t)\,\big],
\end{equation}
with $G_f(t)=\int_{t-\tau}^{t}Y_f\dif\gamma=Y_{fm}\int_{t-\tau}^{t}f\dif\gamma$ and $G_J(t)=\int_{t-\tau}^{t}Y_J\dif\gamma$, i.e.,
\begin{equation}\label{eq:GfGJ}
G_J(t)=\begin{bmatrix}[\Omega_1]&-\int\Omega_2\Omega_3&\int\Omega_2\Omega_3\\ \int\Omega_1\Omega_3&[\Omega_2]&-\int\Omega_1\Omega_3\\ -\int\Omega_1\Omega_2&\int\Omega_1\Omega_2&[\Omega_3]\end{bmatrix},
\end{equation}
where $\int(\cdot)$ denotes $\int_{t-\tau}^{t}(\cdot)\dif\gamma$ and $[\Omega_i]=\Omega_i(t)-\Omega_i(t-\tau)$, so that $\dot\Omega$ is not required.


 We propose the concurrent learning update law
\begin{equation}\label{eq:clr}
\dot{\hat\theta}_r=k_r\Gamma_r\sum_{k=1}^{N_r}G_{k,r}^T\big(P_{k,r}-G_{k,r}\hat\theta_r\big),
\end{equation}
where $k_r\in\R_{>0}$ is a constant adaptation gain, and $\Gamma_r\in\R^{5\times5}$ is the least-squares gain updated using the update law of~\cite{thapa2019concurrent}
\begin{equation}\label{eq:Gammar}
\dot\Gamma_r=\beta_r\Gamma_r-k_r\Gamma_r\frac{\mathscr G_r}{1+\alpha_r\|\mathscr G_r\|}\Gamma_r,
\end{equation}
in which $\alpha_r,\beta_r\in\R_{>0}$ and $\Gamma_r(0)>0$.
Using arguments similar to~\cite{thapa2019concurrent},~\cite{kamalapurkar2017simultaneous} and~\cite[Corollary~4.3.2]{ioannou1996robust}, it can be shown that, provided $\lambda_{\min}\{\Gamma_i^{-1}(0)\}>0$, the gains satisfy
\begin{equation}\label{eq:G_gain}
\underline{\Gamma_i}I\le\Gamma_i(t)\le\overline{\Gamma_i}I,\qquad i\in\{t,r\},
\end{equation}
where $\underline{\Gamma_i}$ and $\overline{\Gamma_i}$ are positive constants.

A history stack, i.e., a set of ordered pairs $\{(P_{k,r},G_{k,r})\}_{k=1}^{N_r}$ such that $P_{k,r}=G_{k,r}\theta_r$, is called \emph{full rank} if there exists a constant $\underline c_r\in\R$ such that
\begin{equation}\label{eq:rankr}
0<\underline c_r<\lambda_{\min}\{\mathscr G_r\},\qquad \mathscr G_r:=\textstyle\sum_{k=1}^{N_r}G_{k,r}^TG_{k,r}\RD{.}
\end{equation}
To select the data points in $\mathscr G_r$, a singular value maximization algorithm can be used~\cite{parikhintegral},~\cite{chowdhary2011singular}. If the condition in~\eqref{eq:rankr} is not satisfied, i.e., the matrix $\mathscr G_r$ is not full rank, then data is added to the history stack until $\underline c_r>0$. Once~\eqref{eq:rankr} is satisfied, then a data point is added to $\mathscr G_r$ only if a predefined amount of time has passed since the last change and if it increases the minimum singular value of $\mathscr G_r$. Therefore, although $\mathscr G_r$ can be discontinuous, it is always piecewise continuous with a lower-bounded dwell time. For more details, we refer interested readers to \cite{parikhintegral}.

Our two ICL based estimators are independent of each other, $\hat m$ does not use the offset, and $\hat\theta_r$ does not use the mass. Unlike~\cite{yang2022parameter, thapa2019concurrent}, no stabilizing term is used, so the estimators do not depend on the tracking errors.

\section{Stability Analysis}\label{sec:stability}
The dynamics~\eqref{eq:dyn1}--\eqref{eq:dyn4} with the proposed control laws~\eqref{eq:Fd}, \eqref{eq:f}, \eqref{eq:M} and the update laws~\eqref{eq:clm}, \eqref{eq:clr} takes the following form:
\begin{align}
\dot e_x&=e_v+R\dot{\hat r},\label{eq:cl_x}\\
m\dot e_v&=-k_xe_x-k_ve_v-\tilde m(\ddot x_d-ge_3)\nonumber\\
&\quad-m\,\tfrac{\mathrm d}{\mathrm dt}\big(R\widehat\Omega\tilde r\big)-\big(I_3-Re_3e_3^TR^T\big)F_d,\label{eq:cl_v}\\
\dot\Psi&=e_R^Te_\Omega,\qquad\dot e_R=C\,e_\Omega,\label{eq:cl_R}\\
J_m\dot e_\Omega&=-k_Re_R-k_\Omega e_\Omega-\tilde J_m\big(R^TR_d\dot\Omega_d-\widehat\Omega R^TR_d\Omega_d\big)\nonumber\\
&\quad-\Omega\times\tilde J_m\Omega+\tilde r\times fe_3,\label{eq:cl_W}\\
\dot{\tilde m}&=-k_t\Gamma_t\mathscr G_t\,\tilde m,\qquad
\dot{\tilde\theta}_r=-k_r\Gamma_r\mathscr G_r\,\tilde\theta_r,\label{eq:cl_est}
\end{align}
where $C=\tfrac12\big(\operatorname{tr}(R^TR_d)I_3-R^TR_d\big)$ with $\|C\|\le1$~\cite{lee2010arxiv}, $\tilde J_m=J_m-\hat J_m$, and $\Omega_d$, $\dot\Omega_d$ are the desired angular velocity and acceleration. 
 Full derivation of the closed-loop system~\eqref{eq:cl_x}--\eqref{eq:cl_W} is given in the Appendix.

The desired equilibrium of~\eqref{eq:cl_x}--\eqref{eq:cl_est} is given by
\begin{equation}\label{eq:Estar}
\mathcal E^*=\big\{e_x=e_v=e_R=e_\Omega=0,\ \tilde\theta_r=0,\ \tilde m=0\big\}.
\end{equation}
Convergence to $\mathcal E^*$ means that the estimated CoM tracks $x_d$ and the attitude tracks $R_d$, and that the estimates of the mass, the planar offset and the inertia converge to their true values. Then, by~\eqref{eq:trueerr}, the true CoM also converges to $x_d$. Parameter convergence relies on the following assumption.

\begin{assume}\label{assum:stage}
There exists a $T^*>\tau$ such that over the time interval $[0,T^*)$, the trajectory of the quadrotor provides enough information for $\mathscr G_t$ and $\mathscr G_r$ to become full rank. \IEEEQED
\end{assume}
A hover window is required for $\mathscr G_t$, and rotations about all three body axes are required during the transient response (over $[0,T^*]$, $T^*>\tau$) for $\mathscr G_r$ to be full rank. The following Lyapunov analysis shows that all signals remain bounded before $T^*$; however, exponential convergence is only guaranteed provided $T^*$ in Assumption~\ref{assum:stage} exists.

\begin{theorem}\label{thm:main}
Let $x_d$ be four times continuously differentiable with bounded derivatives, the control laws~\eqref{eq:Fd}, \eqref{eq:f}, \eqref{eq:M}, with the update laws~\eqref{eq:clm} and~\eqref{eq:clr}, ensure that all signals remain bounded and that the desired equilibrium $\mathcal E^*$ in~\eqref{eq:Estar} is locally exponentially stable. \IEEEQED
\end{theorem}

\begin{IEEEproof}
\emph{Mass estimation:} Consider the positive definite candidate Lyapunov function
\begin{equation}\label{eq:Vm}
V_m=\tfrac12\tilde m^2 .
\end{equation}
Using~\eqref{eq:cl_est} and $\dot{\tilde m}=-\dot{\hat m}$, we obtain
\begin{equation}\label{eq:Vmdot}
\begin{aligned}
\dot V_m&=\tilde m\,\dot{\tilde m}=-\tilde m\,\dot{\hat m}\\
&=-k_t\Gamma_t\mathscr G_t\,\tilde m^2\\
&=-2k_t\Gamma_t\mathscr G_t\,V_m\le0 ,
\end{aligned}
\end{equation}
where we use $\tilde m^2=2V_m$ from~\eqref{eq:Vm}, and  $k_t,\Gamma_t>0$ and $\mathscr G_t=\sum_{k=1}^{N_t}\|G_{k,t}\|^2\ge0$.
During the time interval $[0,T^*)$, $V_m(t)\le V_m(0)$ by~\eqref{eq:Vmdot}, and by~\eqref{eq:Vm},
\begin{equation}\label{eq:mbar}
|\tilde m(t)|\le|\tilde m(0)| ,
\end{equation}
Thus, $\tilde m$ is bounded.
During the time interval $[T^*,\infty)$,  from Assumption~\ref{assum:stage}, $\mathscr G_t\ge\underline c_t$   and $\Gamma_t\ge\underline{\Gamma_t}$ by~\eqref{eq:G_gain}, which further gives us, 
\begin{equation}\label{eq:Vmdot2}
\dot V_m\le-2k_t\underline{\Gamma_t}\,\underline c_t\,V_m .
\end{equation}
By the comparison lemma~\cite[Lemma~3.4]{khalil2002nonlinear}, $V_m(t)\le V_m(T^*)e^{-2k_t\underline{\Gamma_t}\,\underline c_t(t-T^*)}$, and using~\eqref{eq:Vm},
\begin{equation}\label{eq:mexp}
|\tilde m(t)|\le|\tilde m(T^*)|\,e^{-k_t\underline{\Gamma_t}\,\underline c_t(t-T^*)} .
\end{equation}
Hence $\tilde m\to0$ exponentially.

\emph{Offset and inertia estimation:} Consider another positive definite candidate Lyapunov function~\cite{thapa2019concurrent}
\begin{equation}\label{eq:Vth}
V_{\theta_r}=\tfrac12\tilde\theta_r^T\Gamma_r^{-1}\tilde\theta_r .
\end{equation}
Using the identity $\dot\Gamma_r^{-1}=-\Gamma_r^{-1}\dot\Gamma_r\Gamma_r^{-1}$ and~\eqref{eq:Gammar},
\begin{equation}\label{eq:dGinv}
\dot\Gamma_r^{-1}=-\beta_r\Gamma_r^{-1}+k_r\Phi_r,\qquad \Phi_r:=\frac{\mathscr G_r}{1+\alpha_r\|\mathscr G_r\|} .
\end{equation}
Using~\eqref{eq:cl_est} and~\eqref{eq:dGinv}, we obtain
\begin{equation}\label{eq:Vthdot}
\begin{aligned}
\dot V_{\theta_r}&=\tilde\theta_r^T\Gamma_r^{-1}\dot{\tilde\theta}_r+\tfrac12\tilde\theta_r^T\dot\Gamma_r^{-1}\tilde\theta_r\\
&=-k_r\tilde\theta_r^T\mathscr G_r\tilde\theta_r-\tfrac{\beta_r}{2}\tilde\theta_r^T\Gamma_r^{-1}\tilde\theta_r+\tfrac{k_r}{2}\tilde\theta_r^T\Phi_r\tilde\theta_r\\
&\le-\tfrac{k_r}{2}\tilde\theta_r^T\mathscr G_r\tilde\theta_r-\beta_rV_{\theta_r}\le0 ,
\end{aligned}
\end{equation}
where the inequality uses $\Phi_r\le\mathscr G_r$, since $1+\alpha_r\|\mathscr G_r\|\ge1$ and $\mathscr G_r\ge0$.
By~\eqref{eq:G_gain}, $V_{\theta_r}$ is bounded below and above by
\begin{equation}\label{eq:Vthquad}
\frac{1}{2\overline{\Gamma_r}}\|\tilde\theta_r\|^2\le V_{\theta_r}\le\frac{1}{2\underline{\Gamma_r}}\|\tilde\theta_r\|^2 .
\end{equation}
During the time interval $[0,T^*)$, $\dot V_{\theta_r}\le0$, which is negative semi-definite. Hence $V_{\theta_r}(t)\le V_{\theta_r}(0)$, and by~\eqref{eq:Vthquad},
\begin{equation}\label{eq:thrbar}
\|\tilde\theta_r(t)\|\le\sqrt{\overline{\Gamma_r}/\underline{\Gamma_r}}\,\|\tilde\theta_r(0)\| ,
\end{equation}
i.e., $\tilde\theta_r$ is bounded.
During the time interval $[T^*,\infty)$, $\mathscr G_r\ge\underline c_rI$ by Assumption~\ref{assum:stage}, and $\|\tilde\theta_r\|^2\ge2\underline{\Gamma_r}V_{\theta_r}$ by~\eqref{eq:Vthquad}. Using these in~\eqref{eq:Vthdot},
\begin{equation}\label{eq:Vthdot2}
\begin{aligned}
\dot V_{\theta_r}&\le-\tfrac{k_r\underline c_r}{2}\|\tilde\theta_r\|^2-\beta_rV_{\theta_r}\\
&\le-\kappa_rV_{\theta_r},\qquad \kappa_r:=k_r\underline c_r\underline{\Gamma_r}+\beta_r .
\end{aligned}
\end{equation}
By the comparison lemma~\cite[Lemma~3.4]{khalil2002nonlinear}, $V_{\theta_r}(t)\le V_{\theta_r}(T^*)e^{-\kappa_r(t-T^*)}$, and by~\eqref{eq:Vthquad},
\begin{equation}\label{eq:threxp}
\|\tilde\theta_r(t)\|\le\sqrt{\overline{\Gamma_r}/\underline{\Gamma_r}}\,\|\tilde\theta_r(T^*)\|\,e^{-\kappa_r(t-T^*)/2} .
\end{equation}
Hence $\tilde\theta_r\to0$ exponentially.

The analysis of the translational and rotational dynamics below follows~\cite{lee2010arxiv}, with an added perturbation due to the estimation errors; we refer the reader to~\cite{lee2010arxiv} for further details.

\emph{Translational dynamics:}
 Consider the positive definite candidate Lyapunov function of~\cite{lee2010arxiv}
\begin{equation}\label{eq:Vx}
V_x=\tfrac12k_x\|e_x\|^2+\tfrac12m\|e_v\|^2+c_1e_x^Te_v ,
\end{equation}
where $0<c_1<\sqrt{k_xm}$. Using~\eqref{eq:cl_x} and~\eqref{eq:cl_v},
\begin{equation}\label{eq:Vxdot0}
\begin{aligned}
\dot V_x&=k_xe_x^T\dot e_x+e_v^T(m\dot e_v)+c_1\dot e_x^Te_v+\tfrac{c_1}{m}e_x^T(m\dot e_v)\\
&=k_xe_x^T\big(e_v+R\dot{\hat r}\big)+c_1\big(e_v+R\dot{\hat r}\big)^Te_v\\
&\quad+\big(e_v+\tfrac{c_1}{m}e_x\big)^T\Big(-k_xe_x-k_ve_v-\tilde m(\ddot x_d-ge_3)\\
&\qquad-m\,\tfrac{\mathrm d}{\mathrm dt}\big(R\widehat\Omega\tilde r\big)-\big(I_3-Re_3e_3^TR^T\big)F_d\Big)\\
&=-\tfrac{c_1k_x}{m}\|e_x\|^2-(k_v-c_1)\|e_v\|^2-\tfrac{c_1k_v}{m}e_x^Te_v\\
&\quad-\big(e_v+\tfrac{c_1}{m}e_x\big)^T\big(I_3-Re_3e_3^TR^T\big)F_d\\
&\quad-\big(e_v+\tfrac{c_1}{m}e_x\big)^T\Big(\tilde m(\ddot x_d-ge_3)+m\,\tfrac{\mathrm d}{\mathrm dt}\big(R\widehat\Omega\tilde r\big)\Big)\\
&\quad+\big(k_xe_x+c_1e_v\big)^TR\dot{\hat r},
\end{aligned}
\end{equation}
where the last step uses $k_xe_x^Te_v-k_xe_v^Te_x=0$. Let $z_1=[\|e_x\|,\|e_v\|]^T$, $z_2=[\|e_R\|,\|e_\Omega\|]^T$. For the second line, since $F_d=-\|F_d\|R_de_3$, $\|\big(I_3-Re_3e_3^TR^T\big)F_d\|\le\|F_d\|\|e_R\|\le\big(k_x\|e_x\|+k_v\|e_v\|+B\big)\|e_R\|$~\cite{lee2010arxiv}, where $B:=\sup_t\|\hat m(\ddot x_d-ge_3)\|$. On the domain of~\cite[Prop.~2]{lee2010arxiv}, where $\Psi\le\psi_1<1$, $\|e_R\|\le\alpha:=\sqrt{\psi_1(2-\psi_1)}<1$ and $\|e_v\|\le e_{v,\max}$. Using these bounds and $e_x^Te_v\ge-\|e_x\|\|e_v\|$, the~fourth and fifth lines of~\eqref{eq:Vxdot0} are bounded by $-z_1^TW_1z_1+z_1^TW_{12}z_2$~\cite[Prop.~2]{lee2010arxiv}, where
\begin{equation}\label{eq:Wx}
\begin{aligned}
W_1&=\begin{bmatrix}\tfrac{c_1k_x}{m}(1-\alpha)&-\tfrac{c_1k_v}{2m}(1+\alpha)\\-\tfrac{c_1k_v}{2m}(1+\alpha)&k_v(1-\alpha)-c_1\end{bmatrix},\\
W_{12}&=\begin{bmatrix}k_xe_{v,\max}+\tfrac{c_1}{m}B&0\\B&0\end{bmatrix}.
\end{aligned}
\end{equation}
We now get
\begin{equation}\label{eq:Vxdot}
\dot V_x\le-z_1^TW_1z_1+z_1^TW_{12}z_2+b_x\|z_1\|\|\tilde\theta\|
\end{equation}
for some $b_x>0$ that depends on the bounds of $x_d$ and its derivatives, the errors on the domain, and $\overline{\Gamma_r}$.

\emph{Rotational dynamics:} Consider the positive definite candidate Lyapunov function of~\cite{lee2010arxiv}
\begin{equation}\label{eq:VR}
V_R=\tfrac12e_\Omega^TJ_me_\Omega+k_R\Psi+c_2e_R^Te_\Omega ,
\end{equation}
where $0<c_2<\sqrt{k_R\lambda_{\min}(J_m)}$. Using~\eqref{eq:cl_R} and~\eqref{eq:cl_W},
\begin{equation}\label{eq:VRdot0}
\begin{aligned}
\dot V_R&=e_\Omega^T(J_m\dot e_\Omega)+k_R\dot\Psi+c_2\dot e_R^Te_\Omega+c_2e_R^TJ_m^{-1}(J_m\dot e_\Omega)\\
&=-k_\Omega\|e_\Omega\|^2-c_2k_Re_R^TJ_m^{-1}e_R-c_2k_\Omega e_R^TJ_m^{-1}e_\Omega\\
&\quad+c_2e_\Omega^TC^Te_\Omega\\
&\quad+\big(e_\Omega+c_2J_m^{-1}e_R\big)^T\Big(-\tilde J_m\big(R^TR_d\dot\Omega_d-\widehat\Omega R^TR_d\Omega_d\big)\\
&\qquad-\Omega\times\tilde J_m\Omega+\tilde r\times fe_3\Big),
\end{aligned}
\end{equation}
where the last step uses $-k_Re_\Omega^Te_R+k_Re_R^Te_\Omega=0$. Using $\|C\|\le1$, $e_R^TJ_m^{-1}e_R\ge\|e_R\|^2/\lambda_{\max}(J_m)$ and $|e_R^TJ_m^{-1}e_\Omega|\le\|e_R\|\|e_\Omega\|/\lambda_{\min}(J_m)$, the~second and third lines of~\eqref{eq:VRdot0} are bounded by $-z_2^TW_2z_2$, where
\begin{equation}\label{eq:WR}
W_2=\begin{bmatrix}\tfrac{c_2k_R}{\lambda_{\max}(J_m)}&-\tfrac{c_2k_\Omega}{2\lambda_{\min}(J_m)}\\-\tfrac{c_2k_\Omega}{2\lambda_{\min}(J_m)}&k_\Omega-c_2\end{bmatrix} .
\end{equation}
We now get
\begin{equation}\label{eq:VRdot}
\dot V_R\le-z_2^TW_2z_2+b_R\|z_2\|\|\tilde\theta_r\|
\end{equation}
for some $b_R>0$ that depends on the bounds of $x_d$ and its derivatives, the errors on the domain, and $\overline{\Gamma_r}$.

\begin{remark}\label{rem:order}
The mass is identified first, followed by the offset and the inertia. During the time interval $[0,T^*)$, $\dot V_m\le0$ and $\dot V_{\theta_r}\le0$ by~\eqref{eq:Vmdot} and~\eqref{eq:Vthdot}, which are negative semi-definite. It follows from~\cite[Th.~4.8]{khalil2002nonlinear} that $\tilde m$ and $\tilde\theta_r$ are uniformly bounded. However, parameter convergence may not be achieved due to the lack of PE.
During the time interval $[T^*,\infty)$, since $\mathscr G_t$ and $\mathscr G_r$ are full rank, by~\eqref{eq:mexp} and~\eqref{eq:threxp}, $\tilde m\to0$ and $\tilde\theta_r\to0$ exponentially, which further implies that $\hat m\to m$ and $\hat\theta_r\to\theta_r$.
\end{remark}

\emph{Concurrent learning:} Let
\begin{equation}\label{eq:VCL}
V_{CL}=V_m+V_{\theta_r} .
\end{equation}
During the time interval $[T^*,\infty)$, using~\eqref{eq:Vmdot2} and~\eqref{eq:Vthdot2}, we can further simplify $\dot V_{CL}$ to
\begin{equation}\label{eq:VCLdot}
\begin{aligned}
\dot V_{CL}&=\dot V_m+\dot V_{\theta_r}\\
&\le-2k_t\underline{\Gamma_t}\,\underline c_t\,V_m-\kappa_rV_{\theta_r}\\
&\le-\kappa_{CL}V_{CL},\qquad \kappa_{CL}:=\min\{2k_t\underline{\Gamma_t}\,\underline c_t,\ \kappa_r\} .
\end{aligned}
\end{equation}

\emph{Tracking controller:} Let
\begin{equation}\label{eq:VTC}
V_{TC}=V_x+V_R .
\end{equation}
Let $z=[z_1^T,\ z_2^T]^T$ and $W=\left[\begin{smallmatrix}W_1&-\frac12W_{12}\\-\frac12W_{12}^T&W_2\end{smallmatrix}\right]$, where $c_1$ and $c_2$ are chosen as in~\cite[Prop.~2]{lee2010arxiv}, and the gains satisfy $\lambda_{\min}(W_2)>\|W_{12}\|^2/\big(4\lambda_{\min}(W_1)\big)$, so that $W>0$ by the Schur complement. Adding~\eqref{eq:Vxdot} and~\eqref{eq:VRdot} and using $\|\tilde\theta_r\|\le\|\tilde\theta\|$,
\begin{equation}\label{eq:V1dot}
\begin{aligned}
\dot V_{TC}&=\dot V_x+\dot V_R\\
&\le-z^TWz+b\,\|z\|\|\tilde\theta\|\\
&\le-\lambda_{\min}(W)\|z\|^2+b\,\|z\|\|\tilde\theta\| .
\end{aligned}
\end{equation}
\emph{Combined system:} Let
\begin{equation}\label{eq:Vtot}
V=V_{TC}+\rho V_{CL},
\end{equation}
where $\rho>0$, and
\begin{equation}\label{eq:Vdot_def}
\dot V=\dot V_{TC}+\rho\dot V_{CL}.
\end{equation}
During the time interval $[0,T^*)$, $\|\tilde\theta\|\le\bar\theta$ for some $\bar\theta>0$ by Remark~\ref{rem:order}, and by~\eqref{eq:V1dot}, $\dot V_{TC}\le-\lambda_{\min}(W)\|z\|^2+b\bar\theta\|z\|$. It follows from~\cite[Th.~4.18]{khalil2002nonlinear} that $e_x$, $e_v$, $e_R$ and $e_\Omega$ are uniformly ultimately bounded, and $\tilde m$ and $\tilde\theta_r$ are bounded by Remark~\ref{rem:order}. Given that the states are bounded, we further conclude that $\mathscr G_t$ and $\mathscr G_r$ are bounded.

During the time interval $[T^*,\infty)$, using~\eqref{eq:V1dot} and~\eqref{eq:VCLdot} in~\eqref{eq:Vdot_def}, we obtain
\begin{equation}\label{eq:Vfinal}
\dot V\le-\lambda_{\min}(W)\|z\|^2+b\,\|z\|\|\tilde\theta\|-\rho\kappa_{CL}V_{CL} .
\end{equation}
Since $V_{CL}\ge\underline v_{CL}\|\tilde\theta\|^2$ with $\underline v_{CL}=\tfrac12\min\{1,1/\overline{\Gamma_r}\}$ by~\eqref{eq:G_gain},~\eqref{eq:Vfinal} reduces to
\begin{equation}\label{eq:VdotQ}
\dot V\le-\begin{bmatrix}\|z\|\\ \|\tilde\theta\|\end{bmatrix}^T
\underbrace{\begin{bmatrix}\lambda_{\min}(W)&-\tfrac b2\\-\tfrac b2&\rho\kappa_{CL}\underline v_{CL}\end{bmatrix}}_{Q}
\begin{bmatrix}\|z\|\\ \|\tilde\theta\|\end{bmatrix} .
\end{equation}
The matrix $Q$ is positive definite for
\begin{equation}\label{eq:rho}
\rho>\frac{b^2}{4\lambda_{\min}(W)\kappa_{CL}\underline v_{CL}} ,
\end{equation}
so that $\dot V\le-\lambda_{\min}(Q)\big(\|z\|^2+\|\tilde\theta\|^2\big)$, which is negative definite. Using $\underline v\|z\|^2\le V_{TC}\le\bar v\|z\|^2$ for some $\underline v,\bar v>0$~\cite{lee2010arxiv} and $V_{CL}\le\bar v_{CL}\|\tilde\theta\|^2$ with $\bar v_{CL}=\tfrac12\max\{1,1/\underline{\Gamma_r}\}$ by~\eqref{eq:G_gain}, we obtain
\begin{equation}\label{eq:Vquad}
k_1\big(\|z\|^2+\|\tilde\theta\|^2\big)\le V\le k_2\big(\|z\|^2+\|\tilde\theta\|^2\big),
\end{equation}
where $k_1=\min\{\underline v,\rho\underline v_{CL}\}$ and $k_2=\max\{\bar v,\rho\bar v_{CL}\}$. By introducing a dwell time in the singular value maximization algorithm, it can be easily ensured that the points of discontinuity of $\mathscr G_t$ and $\mathscr G_r$ do not accumulate. We apply~\cite[Th.~4.10]{khalil2002nonlinear} and conclude that $e_x,e_v,e_R,e_\Omega,\tilde m,\tilde\theta_r\to0$ exponentially, i.e., $\mathcal E^*$ is locally exponentially stable, which further implies that $R\to R_d$, $\hat m\to m$ and $\hat\theta_r\to\theta_r$, and, by~\eqref{eq:trueerr}, $x_m\to x_d$.
\end{IEEEproof}

\section{Numerical Simulations}\label{sec:sim}
\begin{figure*}[t]
\centering
\includegraphics[width=\textwidth]{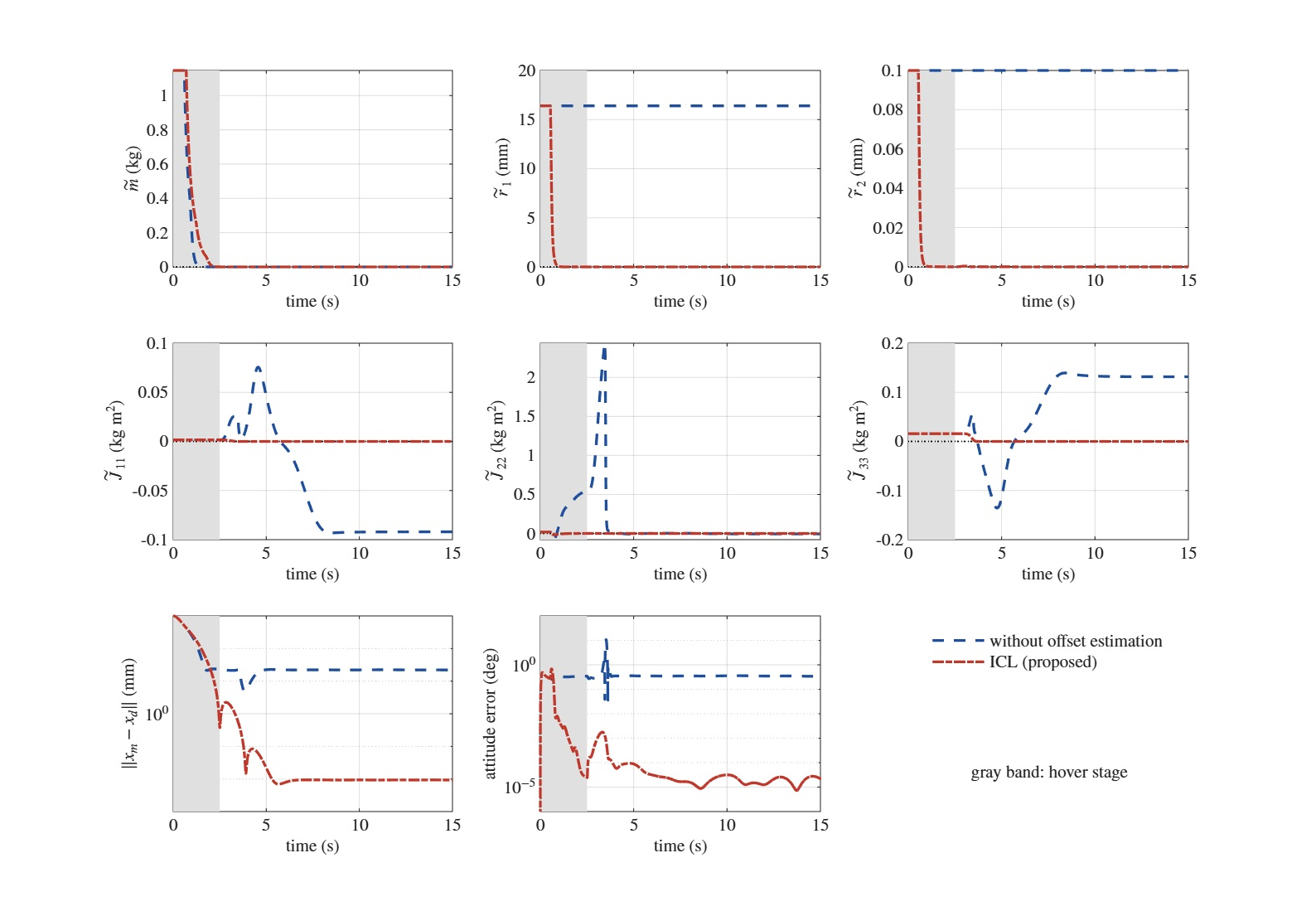}
\caption{Parameter estimation errors $\tilde\theta=\theta-\hat\theta$, and position and attitude errors of the proposed ICL controller with offset estimation  and without offset estimation (blue, dashed). The vehicle hovers until $2.5$~s (gray band).}
\label{fig:est}
\end{figure*}
\begin{figure*}[t]
\centering
\includegraphics[width=\textwidth]{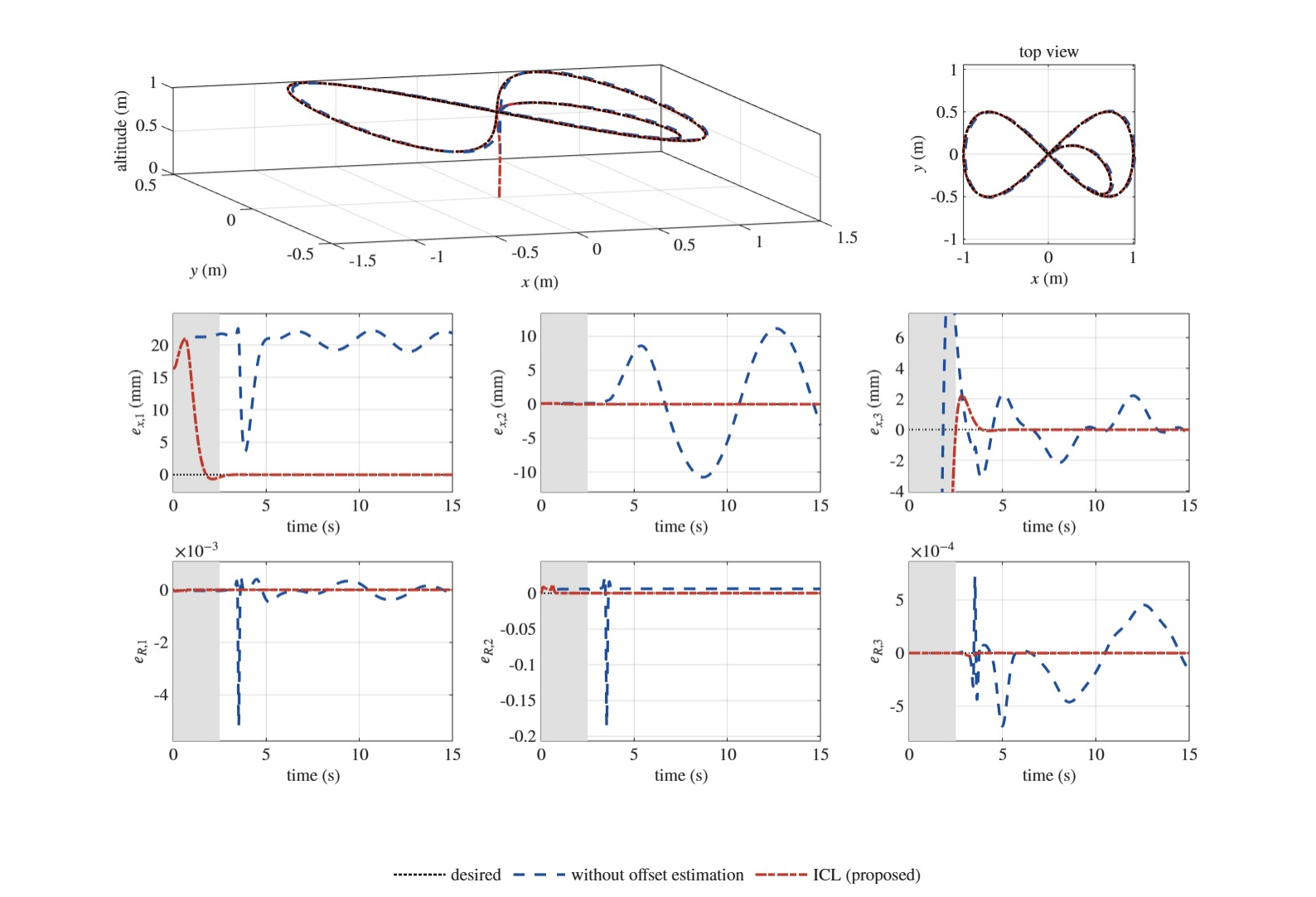}
\caption{3D trajectory tracking of the proposed ICL controller with offset estimation and without offset estimation. Left: true CoM trajectory $x_m$ and desired trajectory $x_d$. Right: top view for $0\le t\le15$~s.}
\label{fig:traj}
\end{figure*}
We validate the effectiveness of the proposed algorithms in simulation, without and with offset estimation\footnote{The MATLAB code of this letter is available at \url{https://github.com/sandeshthapa/Concurrent_Learning_Control}.}. The offset-unaware controller uses $\hat r\equiv0$ and estimates only the mass and the inertia. We use the following desired trajectory and heading:
\begin{equation}\label{eq:fig8}
\begin{aligned}
x_d(t)&=[0,\ 0,\ -1]^T,\quad b_{1d}(t)=e_1,\quad 0\le t<2.5,\\
x_d(t)&=\begin{bmatrix}\big(1-e^{-0.1(t-2.5)^3}\big)\sin\big(\tfrac{\pi}{4}(t-2.5)\big)\\[2pt] 0.5\big(1-e^{-0.1(t-2.5)^3}\big)\sin\big(\tfrac{\pi}{2}(t-2.5)\big)\\[2pt] -1\end{bmatrix}\!,\ t\ge2.5,\\
b_{1d}(t)&=\begin{bmatrix}\cos\phi_d(t)\\ \sin\phi_d(t)\\ 0\end{bmatrix},\quad t\ge2.5,\\
\phi_d(t)&=0.5\big(1-e^{-0.1(t-2.5)^3}\big)\sin\big(\tfrac{\pi}{4}(t-2.5)\big),
\end{aligned}
\end{equation}
where $x_d$ is in meters and $\phi_d$ in rad.

The true mass of the quadrotor is $m=3.746$~kg, with the offset at $r_m=[1.64,\ 0.01,\ 0]^T$~cm and the inertia $J_m=\diag(0.1015,\ 0.1181,\ 0.2158)$~kg\,m$^2$. Our initial estimates are $\hat m(0)=2.6$~kg, $\hat r(0)=0$ and $\hat J_m(0)=\diag(0.1,\ 0.1,\ 0.2)$~kg\,m$^2$. The gains are set to be $k_x=40$, $k_v=18$, $k_R=100$, $k_\Omega=5.1$, $k_r=k_t=10$, $\beta_r=\beta_t=1$, $\alpha_r=3\times10^{-4}$, $\alpha_t=4.5\times10^{-3}$, $\Gamma_r(0)=\diag(10^{-3}I_2,\ 0.5I_3)$ and $\Gamma_t(0)=10^{-2}$, with $\tau=0.5$~s, windows recorded every $0.02$~s and $N_r=N_t=400$.

Fig.~\ref{fig:est} shows that the mass is learned within a few seconds. Our offset-aware controller is also able to learn the offset and all three inertias. The offset-unaware controller is able to learn the mass quickly as well. However, to compensate for the offset, the offset-unaware controller has constant errors in all three inertia estimates, which are not present in our offset-aware controller. The largest errors are in $\hat J_{11}$, which is approximately $90\,\%$ too large, and $\hat J_{33}$, which is approximately $60\,\%$ too small, while $\hat J_{22}$ is approximately $26\,\%$ too large. In contrast, with our offset-aware controller, the errors in $\hat J_{11}$, $\hat J_{22}$ and $\hat J_{33}$ are within $1\,\%$ after $3.0$, $2.2$ and $3.5$~s, respectively, and converge to zero. Our CoM error also remains bounded, and stays below $1$~mm after $3.5$~s.

Fig.~\ref{fig:traj} shows a detailed view of the tracking of the two approaches.

\section{Conclusion}\label{sec:conc}
In this letter, we considered a quadrotor whose CoM is offset from its geometric center. We developed a concurrent learning-based estimation and control algorithm that estimates the planar CoM offset, the moments of inertia, and the mass of the vehicle with its payload. We learn the these parameters in stages: the mass and the offset in hover, and the inertia with a 3D trajectory. To the best of our knowledge, no concurrent learning algorithms have been developed to estimate the CoM offset together with the inertia and the mass. Future work will focus on experimental validation of the algorithms. Dynamic changes and purging will also be considered to handle multiple time-varying moments of inertia and other parameters. Drag and wind effects similar to~\cite{chen2022model} will also be considered.

\appendices
\section{Derivation of the Closed-Loop Error Dynamics}\label{app:cl}
\emph{Translational dynamics.} Plugging~\eqref{eq:xhat} into~\eqref{eq:errors}, we get
\begin{align}
e_x&=x_o+R\hat r-x_d,\label{eq:r_ex}\\
e_v&=v_o+R\widehat\Omega\hat r-\dot x_d .\label{eq:r_ev}
\end{align}
Plugging~\eqref{eq:vmdot} into the translational dynamics~\eqref{eq:dyn2}, we get
\begin{align}
m\dot v_m&=mge_3-fRe_3,\label{eq:r_dyn}\\
m\Big(\dot v_o+\tfrac{\mathrm d}{\mathrm dt}\big(R\widehat\Omega\big)r_m\Big)&=mge_3-fRe_3,\label{eq:r_plug}\\
m\dot v_o&=mge_3-fRe_3\nonumber\\
&\quad-m\,\tfrac{\mathrm d}{\mathrm dt}\big(R\widehat\Omega\big)r_m .\label{eq:r_mvo}
\end{align}
Differentiating $e_x$ and using $\dot x_o=v_o$ and $\dot R=R\widehat\Omega$,
\begin{align}
\dot e_x&=\dot x_o+\dot R\hat r+R\dot{\hat r}-\dot x_d\label{eq:r_dex1}\\
&=v_o+R\widehat\Omega\hat r+R\dot{\hat r}-\dot x_d\label{eq:r_dex2}\\
&=e_v+R\dot{\hat r}.\label{eq:r_dex}
\end{align}
Differentiating $e_v$, multiplying by $m$, and plugging in $m\dot v_o$ from above,
\begin{align}
m\dot e_v&=m\dot v_o+m\,\tfrac{\mathrm d}{\mathrm dt}\big(R\widehat\Omega\big)\hat r+mR\widehat\Omega\dot{\hat r}-m\ddot x_d\label{eq:r_mdev1}\\
&=mge_3-fRe_3-m\ddot x_d-m\,\tfrac{\mathrm d}{\mathrm dt}\big(R\widehat\Omega\big)r_m\nonumber\\
&\quad+m\,\tfrac{\mathrm d}{\mathrm dt}\big(R\widehat\Omega\big)\hat r+mR\widehat\Omega\dot{\hat r}\label{eq:r_mdev2}\\
&=mge_3-fRe_3-m\ddot x_d-m\,\tfrac{\mathrm d}{\mathrm dt}\big(R\widehat\Omega\big)\tilde r-mR\widehat\Omega\dot{\tilde r}\label{eq:r_mdev3}\\
&=mge_3-fRe_3-m\ddot x_d-m\,\tfrac{\mathrm d}{\mathrm dt}\big(R\widehat\Omega\tilde r\big),\label{eq:r_mdev}
\end{align}
where~\eqref{eq:r_mdev3} uses $\tilde r=r_m-\hat r$ and $\dot{\tilde r}=-\dot{\hat r}$, since $r_m$ is constant, and~\eqref{eq:r_mdev} uses the product rule.
Plugging $f=-F_d^TRe_3$ from~\eqref{eq:f} into~\eqref{eq:r_mdev}, we get
\begin{align}
m\dot e_v&=mge_3+\big(F_d^TRe_3\big)Re_3-m\ddot x_d-m\,\tfrac{\mathrm d}{\mathrm dt}\big(R\widehat\Omega\tilde r\big)\label{eq:r_f1}\\
&=mge_3+Re_3e_3^TR^TF_d-m\ddot x_d-m\,\tfrac{\mathrm d}{\mathrm dt}\big(R\widehat\Omega\tilde r\big),\label{eq:r_f2}
\end{align}
where~\eqref{eq:r_f2} uses that $F_d^TRe_3=e_3^TR^TF_d$ is a scalar. Plugging $F_d=-k_xe_x-k_ve_v+\hat m(\ddot x_d-ge_3)$ from~\eqref{eq:Fd} into~\eqref{eq:r_f2}, we get
\begin{align}
m\dot e_v&=mge_3-m\ddot x_d-m\,\tfrac{\mathrm d}{\mathrm dt}\big(R\widehat\Omega\tilde r\big)\nonumber\\
&\quad+Re_3e_3^TR^T\big(-k_xe_x-k_ve_v+\hat m(\ddot x_d-ge_3)\big).\label{eq:r_f3}
\end{align}
Writing $Re_3e_3^TR^T=I_3-\big(I_3-Re_3e_3^TR^T\big)$ in~\eqref{eq:r_f3}, and using~\eqref{eq:Fd} for the second part, we get
\begin{align}
m\dot e_v&=mge_3-m\ddot x_d+\hat m(\ddot x_d-ge_3)-k_xe_x-k_ve_v\nonumber\\
&\quad-m\,\tfrac{\mathrm d}{\mathrm dt}\big(R\widehat\Omega\tilde r\big)-\big(I_3-Re_3e_3^TR^T\big)F_d.\label{eq:r_f4}
\end{align}
Since $mge_3-m\ddot x_d+\hat m(\ddot x_d-ge_3)=-(m-\hat m)(\ddot x_d-ge_3)=-\tilde m(\ddot x_d-ge_3)$,~\eqref{eq:r_f4} reduces to
\begin{align}
m\dot e_v&=-k_xe_x-k_ve_v-\tilde m(\ddot x_d-ge_3)\nonumber\\
&\quad-m\,\tfrac{\mathrm d}{\mathrm dt}\big(R\widehat\Omega\tilde r\big)-\big(I_3-Re_3e_3^TR^T\big)F_d.\label{eq:r_f5}
\end{align}
The last term vanishes when $Re_3$ is parallel to $F_d$, i.e., when the attitude tracks $R_d$.

\emph{Rotational dynamics.} From the rotational dynamics~\eqref{eq:dyn4}, we get
\begin{equation}
J_m\dot\Omega=M+r_m\times fe_3-\Omega\times J_m\Omega .\label{eq:r_rot1}
\end{equation}
Plugging $M$ from~\eqref{eq:M} into~\eqref{eq:r_rot1}, we get
\begin{align}
J_m\dot\Omega&=-k_Re_R-k_\Omega e_\Omega+\hat J_m\big(R^TR_d\dot\Omega_d-\widehat\Omega R^TR_d\Omega_d\big)\nonumber\\
&\quad+\Omega\times\hat J_m\Omega-\hat r\times fe_3+r_m\times fe_3\nonumber\\
&\quad-\Omega\times J_m\Omega .\label{eq:r_rot2}
\end{align}
Differentiating $e_\Omega=\Omega-R^TR_d\Omega_d$ and using $\dot R=R\widehat\Omega$ and $\dot R_d=R_d\widehat\Omega_d$,
\begin{align}
\dot e_\Omega&=\dot\Omega-\dot R^TR_d\Omega_d-R^T\dot R_d\Omega_d-R^TR_d\dot\Omega_d\label{eq:r_deW1}\\
&=\dot\Omega+\widehat\Omega R^TR_d\Omega_d-R^TR_d\widehat\Omega_d\Omega_d-R^TR_d\dot\Omega_d\label{eq:r_deW2}\\
&=\dot\Omega-\big(R^TR_d\dot\Omega_d-\widehat\Omega R^TR_d\Omega_d\big),\label{eq:r_deW}
\end{align}
where~\eqref{eq:r_deW2} uses $\widehat\Omega^{\,T}=-\widehat\Omega$, and~\eqref{eq:r_deW} uses $\widehat\Omega_d\Omega_d=\Omega_d\times\Omega_d=0$. Multiplying~\eqref{eq:r_deW} by $J_m$ and plugging in $J_m\dot\Omega$ from~\eqref{eq:r_rot2},
\begin{align}
J_m\dot e_\Omega&=J_m\dot\Omega-J_m\big(R^TR_d\dot\Omega_d-\widehat\Omega R^TR_d\Omega_d\big)\label{eq:r_rot3}\\
&=-k_Re_R-k_\Omega e_\Omega\nonumber\\
&\quad+\big(\hat J_m-J_m\big)\big(R^TR_d\dot\Omega_d-\widehat\Omega R^TR_d\Omega_d\big)\nonumber\\
&\quad+\Omega\times\hat J_m\Omega-\Omega\times J_m\Omega\nonumber\\
&\quad+r_m\times fe_3-\hat r\times fe_3 .\label{eq:r_rot4}
\end{align}
Since $J_m-\hat J_m=\tilde J_m$ and $r_m-\hat r=\tilde r$,~\eqref{eq:r_rot4} reduces to
\begin{align}
J_m\dot e_\Omega&=-k_Re_R-k_\Omega e_\Omega-\tilde J_m\big(R^TR_d\dot\Omega_d-\widehat\Omega R^TR_d\Omega_d\big)\nonumber\\
&\quad-\Omega\times\tilde J_m\Omega+\tilde r\times fe_3 .\label{eq:r_rot}
\end{align}
Here $\Omega_d$ and $\dot\Omega_d$ are the angular velocity and acceleration of $R_d$.

\bibliographystyle{IEEEtran}
\bibliography{IEEEabrv,mybib}
\end{document}